\documentclass[%
 reprint,
 amsmath,amssymb,
 aps,
]{revtex4-1}
\usepackage[dvipdfmx]{graphicx}
\usepackage{dcolumn}
\usepackage{bm}
\usepackage{xcolor}

\begin{document}


\title{The Polymer Physics of Kinetoplast DNA as a Polymerised Membrane}

\author{Takahiro Sakaue}
\email{sakaue@phys.aoyama.ac.jp}
 \affiliation{%
 Department of Physical Sciences, Aoyama Gakuin University, 5-10-1 Fuchinobe, Chuo-ku, Sagamihara, Japan
}%
\author{Davide Michieletto}%
 \email{davide.michieletto@ed.ac.uk}
\affiliation{School of Physics and Astronomy, University of Edinburgh, Edinburgh Eh9 3FD, Scotland}
\affiliation{MRC Human Genetics Unit, Institute of Genetics and Cancer, University of Edinburgh, Scotland}
\affiliation{International Institute for Sustainability with Knotted Chiral Meta Matter (WPI-SKCM$^2$), Hiroshima University, Higashi-Hiroshima, Hiroshima 739-8526, Japan}

\date{\today}

\begin{abstract}
We analyze the conformational and dynamical properties of the kinetoplast DNA (kDNA), a massive sheet-like structure made from thousands of circular DNA molecules, found in the mitochondrion of certain parasites. The connectivity between circular DNA molecules is achieved by topological linking, hence, the kDNA may be regarded as a naturally occurring two-dimensional version of Olympic gels, whose physical properties are yet to be understood. We propose that the basic aspects in the large scale behaviors of kDNA could be described by the physics of polymerized membrane. Our analysis indicates the relevance of the hydrodynamic interactions in the dynamics of kDNA in aqueous solution. We demonstrate that the predicted dynamical scaling scenario captures various experimental data recently obtained from {\it in vitro} imaging experiments in a unified manner. We also provide an estimate for the in-plane elastic modulus of kDNA, whose magnitude agrees well with recent measurements.
\end{abstract}

\maketitle

\section{Introduction}
Topological constraints are fundamental in determining the equilibrium and dynamical properties of polymer systems~\cite{Rubinstein_book}. While entanglements are extensively studied in the context of polymer melts~\cite{Doi_book}, topologically linked polymer systems have received far less attention likely because more challenging to synthesise~\cite{deGennes_book}. One of the best examples of a catenated polymer structure abundant in nature is the mitochondrial genome of Trypanosomes, known as kinetoplast DNA (kDNA)~\cite{Simpson1979,Chen1995,Michieletto2025nar}. This structure consists of thousands of covalently closed DNA loops organized into a highly connected network through topological linkages rather than chemical crosslinks~\cite{Michieletto2025nar,Tubiana2024}.

In \textit{Crithidia fasciculata}, the network comprises $\sim$5000 minicircles ($\sim$2.5 kb) and several dozen maxicircles ($>$10 kb), forming a two-dimensional arrangement confined within the mitochondrion~\cite{Chen1995}. Each minicircle behaves as a semiflexible ring with contour length on the order of 6 - 16 persistence lengths. Each minicircle is linked to other three minicircles on average~\cite{Chen1995,He2023} placing it near the percolation threshold where the system transitions from disconnected clusters to a fully connected component~\cite{Michieletto2014kdna}. This critical connectivity imparts the network with emergent properties analogous to those predicted for theoretical ``Olympic gels''~\cite{deGennes_book,Michieletto2025nar}, where topology, rather than covalent bonding, governs elasticity and connectivity.

Recent \textit{in vitro} imaging experiments~\cite{Klotz_2019,He2023,Ramakrishnan_2025,Pyne2024kDNA} and simulations~\cite{He2023,Ramakrishnan_2025} have revealed that kDNA behaves as a two-dimensional polymer membrane with non-zero mean curvature -- likely induced by the preferential location of maxicircles at the periphery of the network and/or by a hyper-linked population of minicircles~\cite{Klotz_2019,Ramakrishnan_2025}. 
In Refs~\cite{Klotz_2019}-\cite{Ramakrishnan_2025}, several characteristic time scales, ranging from sub-second to several minutes, have been identified in the dynamics of individual kDNAs through the direct observation of fluorescently labeled kDNAs in aqueous solution. In addition, in Refs.~\cite{He2023} and ~\cite{Ramakrishnan_2025} it was also estimated that the kDNA has a remarkably low in-plane stiffness, about $10^6$ smaller than lipid membranes due to the loose arrangement of minicircles. These features enable large-scale shape transformations and confer mechanical responses markedly different from those of chemically crosslinked gels or lipid bilayers~\cite{JANSHOFF20152977}. Furthermore, the connectivity of kDNA can be actively regulated by enzymatic strand-passing reactions, providing a natural platform to investigate dynamic topological remodeling~\cite{Kim2013a,Krajina2018} -- a process with no synthetic analogue to date.

Thus, kDNA is a unique model for exploring the physics of catenated networks, critical percolation, and topology-driven elasticity.
In this paper, we draw an analogy between kDNA and two-dimensional polymerised membrane. We discover that by accounting for the scaling of the individual network components, we can predict three characteristic time scales, the rotational relaxation time and two distinct internal relaxation times, relevant to the kDNA dynamics. Based on these knowledge, we construct a dynamical scaling scenario for the kDNA dynamics and compare it with a recent measurement from imaging fluorescently labelled quantum dots embedded in the kDNA~\cite{Ramakrishnan_2025}. The analogy also enables us to obtain an estimate of the in-plane elastic modulus of kDNAs, which suggests the anomalous elasticity of the polymerized membrane as a relevant factor for their soft responses. 

Overall, this work contributes to a quantitative understanding of two-dimensional topological materials and thereby not only addresses fundamental questions in soft matter physics but also offers blueprints for engineering synthetic Olympic gels and catenated materials with tunable mechanical and transport properties.

\section{Flory type theory on Kinetoplast DNA}
We first consider the global conformation of kDNA, which is made from $ M \simeq N^2$ rings interlinked into a two-dimensional sheet. Here, for simplicity, we neglect the binary nature of kDNA, assuming that component rings are mono-disperse in size. Each ring, i.e., minicircle, is made from $g$ monomers of size $b$ such that the ring size is $\xi = b g^{\nu}$ with the exponent $\nu$ characterizing the conformation of the rings. This naturally leads to a view that large scale behaviors of kDNA could be modeled as a polymerized membrane, also called as a sheet, where rings are identified as coarse grained monomers. 
Our aim in this section is to set up a description of kDNA based on a simple polymer physics approach, that is the analysis of Flory-type free energy suitably generalized to the network with two-dimensional connectivity.

\paragraph{Historical perspective}
To begin with, we briefly recall that a naive extension of the Flory-type free energy to polymerized membranes results in a wrong prediction~\cite{Kantor_1986}.
 For a membrane composed of $M \simeq N^2$ rings, where $N$ stands for the linear dimension along the two-dimensional connectivity, we may write the free energy as
\begin{eqnarray}
\frac{F(R)}{k_BT} \simeq \frac{R^2}{\xi^2} + v_{\xi} \frac{M^2}{R^3}
\end{eqnarray}
where $R$ is the overall size of the membrane and $v_{\xi} \simeq \xi^3$ is the excluded-volume of each monomer. The first term represents the entropic elasticity of the network, thus favors the shrinkage, while the second term accounting for two-body repulsive interactions counteracts it~\footnote{Note that the elastic free energy for $D$ -dimensional manifolds is evaluated as $\sim k_BT R^2/(\xi^2 N^{2-D})$. The familiar result for linear polymers is recovered for $D=1$, while $D=2$ corresponds to the sheets with two-dimensional connectivity~\cite{Kantor_1986,Kardar_1987}. }.
Minimizing $F(R)$ with respect to $R$, one finds the equilibrium size of the membrane
\begin{eqnarray}
R \simeq (v_{\xi} \xi^2 )^{1/5} M^{2/5} \simeq \xi N^{4/5}
\label{F_R_o}
\end{eqnarray}
For large sheet $N \gg 1$, the above result $R \sim N^{4/5} \ll N$ indicates that the large sheet is much shrunken compared to its flat state, thus, said to be "crumpled".
However, past intensive studies, including large scale numerical simulations, have shown that the polymerized membranes assume a flat conformation in good solvent conditions~\cite{Abraham1989,Kantor1993,Bowick2001,Mizuochi_2014}. The fact that the polymerized membrane does not crumple in three-dimensional space might be received with surprise. It is, however, understood as a consequence of a geometrical constraint in two-dimensional connectivity. Indeed, assuming that the sheet is effectively inextensible, Gauss's Theorema Egregium states that conformations of the initially flat sheet preserve zero Gaussian curvature in the course of fluctuations.

\paragraph{Modified free energy}
Fluorescence microscopy image of kDNA~\cite{Klotz_2019} may give an impression that the kDNA looks crumpled in solution. One may expect that the above description of the crumpled sheet could be applied to the kDNA. However, more recent simulations~\cite{He2023,Ramakrishnan_2025} have shown that a model kDNA with interlinked minicircles, hence, mono-disperse rings, is indeed flat with zero mean curvature. This is consistent with the conformational behavior of polymerized membranes. 

\begin{figure}[t]
	\centering
\includegraphics[width=0.5\textwidth]{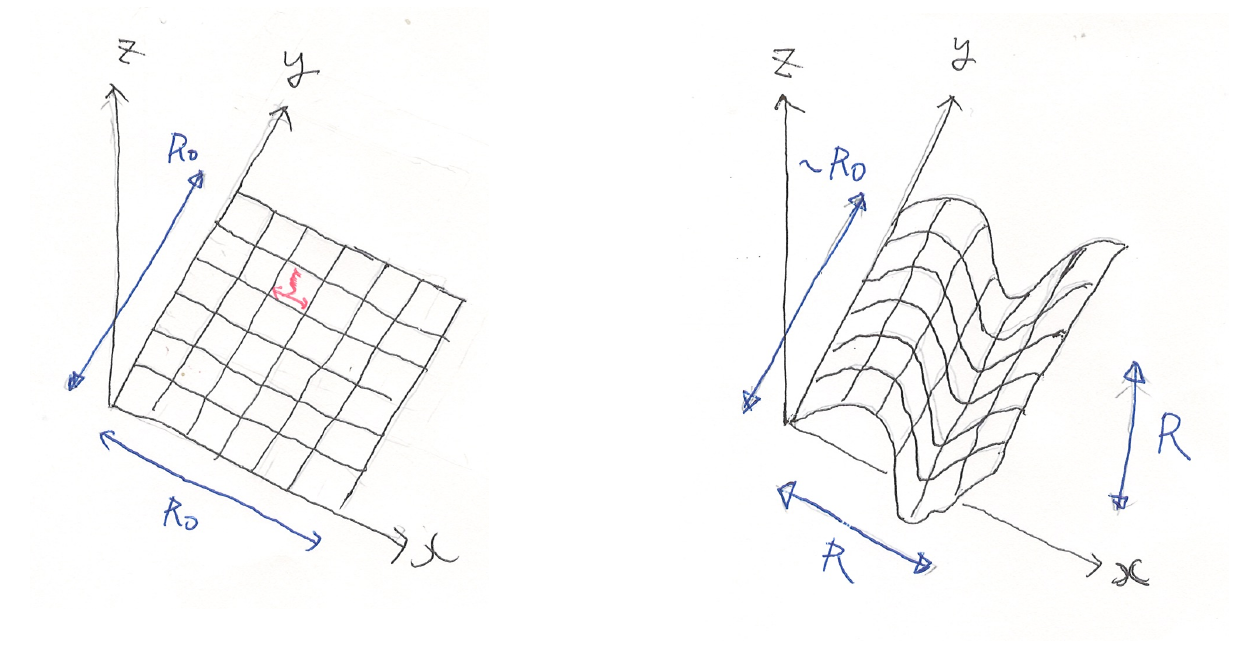}
	\caption{One-dimensional deformation of a sheet creating no Gaussian curvature. Grand state (left) and the deformed state (right).}
	\label{Fig1}
			\vspace{-0.4 cm}
\end{figure}

Given the construction of kDNA network, it is likely that the sheet is locally amenable to in-plane stretching and compression. However, the permanent linkage with neighboring rings might suppress such in-plane freedoms on scale larger than $\xi$, where the network is viewed as effectively inextensible.
The original free energy given by Eq.~(\ref{F_R_o}) allowing arbitrary deformations is inappropriate in this respect. However, one-dimensional fluctuations meet the zero Gaussian curvature constraint and thus are still possible, see Fig.~\ref{Fig1}.

With this constraint in mind, we may consider the following Flory-type free energy;
\begin{eqnarray}
\frac{F(R)}{k_BT} \simeq \frac{R^2}{\xi^2 N} + v_{\xi} \frac{M^2}{R_0 R^2}
\label{F_R_1}
\end{eqnarray}
where $R_0 \simeq \xi N$ represents the undeformed size of the sheet along, say, $y$-direction. Minimizing free energy~(\ref{F_R_1}), we indeed obtain the flat state scaling in thermal equilibrium, i.e.
\begin{eqnarray}
R \simeq \xi N \, .
\label{R_flat}
\end{eqnarray}
The rotational invariance implies that exchanging $x$ and $y$ directions results in the same conclusion and the real conformation would be seen as the combination (or competition) of these deformation modes.

We have seen that the introduction of the geometrical constraint remedies the Flory-type theory to yield the correct scaling in a relatively simple manner. A more precise description requires how the Gaussian curvature that is created by the competition of the two deformation modes mentioned above suppresses the height fluctuation, see Sec.~\ref{PM_kDNA}. 

\paragraph{Dynamics}
The flat state scaling eq.~\eqref{R_flat} has consequences on the dynamics. Here we discuss one aspect.
Assuming the kDNA sheet, the diffusivity of its center of mass is
\begin{eqnarray}
D_{cm}\simeq \left\{
\begin{array}{ll}
\frac{k_BT}{\eta_0 b g M} & (\rm{Rouse})\\
\frac{k_BT}{\eta_0 R} & (\rm{Zimm})
\end{array}
\right.
\label{D_cm}
\end{eqnarray}\\
where $\eta_0$ is the solvent viscosity, and we consider two scenarios Rouse and Zimm dynamics~\cite{deGennes_book,Doi_book,Rubinstein_book}. 
From the relation $D_{cm} \tau_R \simeq R^2$, we then get the time scales
\begin{eqnarray}
\tau_R \simeq \left\{
\begin{array}{ll}
\tau_{\xi}M^{2} \simeq  \tau_{\xi}N^{4}  &(\rm{Rouse})\\
\tau_{\xi}M^{3/2}\simeq \tau_{\xi}N^{3} & (\rm{Zimm})
\label{tau_R_1}
\end{array}
\right.
\end{eqnarray}
where $\tau_{\xi}$ is the ``monomer" time scale of the sheet, i.e., the basic time scale of the sheet, which is set by the relaxation time of each ring
\begin{eqnarray}
\tau_{\xi}= \left\{
\begin{array}{ll}
\frac{\eta_0 b g \xi^2}{k_BT} \simeq \tau_b g^{1+2\nu} & (\rm{Rouse})\\
\frac{\eta_0 \xi^3}{k_BT} \simeq \tau_b g^{3\nu} & (\rm{Zimm})
\end{array}
\right. \label{tau_xi}
\end{eqnarray}
where we introduce the time scale $\tau_b \simeq \eta_0 b^3/k_BT$ of bead (monomer of each ring). The above diffusion time scale $\tau_R$ is expected to have the same scaling as the rotational relaxation time of the sheet~\cite{Mizuochi_2014}. 
If the sheet was isotropically crumpled, the relaxation time of rotation and that of internal deformations have the same scaling structure (given by Eq.~(\ref{t_r_crumpled}) in appendix) just as for flexible polymers. For flat sheets, however, there may naturally arise an anisotropy with distinct in-plane and out-of-plane deformation modes, and these relaxation dynamics would be decoupled from the rotational dynamics of the sheet. We will discuss these aspects in the following section.

\section{Phonon and undulation modes in Kinetoplast DNA}
\label{PM_kDNA}
The sheet in the flat state is not necessarily completely flat. The precise definition of the flat phase is the presence of long-range order in the vector normal to the sheet. The flat sheet is thus not smooth, but rough in the surface normal direction, and this degree of roughening would be related to the elasticity and dynamics of the sheet.

Given the sheet takes the flat state, one can parameterize the conformation of the sheet as ${\vec r}({\vec x}) = (x + u_x({\vec x}), y + u_y({\vec x}), u_{\parallel}({\vec x}))$, where ${\vec x}=(x_1, x_2)$ labels the internal position of ``monomers" within the sheet, and the displacement from the undeformed state is decomposed into the out-of-plane undulation mode $u_{\parallel}({\vec x})$ and the in-plane phonon mode ${\vec u}_{\perp} = (u_x({\vec x}), u_y({\vec x}))$~\cite{Nelson_1987,Nelson_1987_2}.
In our case of kDNA, the monomer corresponds to a ring of size $\xi$, and the continuum description would hold on a scale larger than $\xi$.

Since bending a sheet accompanies its stretching, this geometrical constraint makes the out-of-plane displacement coupled to the in-plane stretching elastic energy.  
Nelson and Peliti showed that this energy, after integrating out the phonon degrees of freedom, can be expressed as interaction between distant Gaussian curvatures, i.e., Gaussian curvature acts as a source for excess energy~\cite{Nelson_1987}. The undulations with nonzero Gaussian curvature will therefore be suppressed in accordance to the consideration leading to Eq.~(\ref{F_R_1}). In addition, this coupling leads to the renormalization of the bending and elastic moduli~\cite{Nelson_1987,Aronovitz1988,Chaikin_book}.
In the following, we review the conformational and dynamical properties of the polymerized membrane in flat phase, and apply these results to rationalize the experimentally observed unique behaviors of kDNA. 

\paragraph{Conformation}
The long-wavelength behaviors of the polymerized membrane could be described by the following effective free energy functional~\cite{Nelson_1987_2,Chaikin_book}
\begin{eqnarray}
{\mathcal H} &=& \frac{1}{2} \int \frac{d^2 q}{(2 \pi)^2} \bigg[ \kappa (q) q^4 |u_{\parallel}({\vec q})|^2  \nonumber \\
&+& \mu(q) q^2 |{\vec u}_{\perp}({\vec q})|^2 + \{\mu(q) + \lambda (q)\}  |{\vec q} \cdot {\vec u}_{\perp}({\vec q})|^2 \bigg]
\label{f_eff}
\end{eqnarray}
in terms of Fourier modes $u_{\parallel}({\vec q}) = \int_{-\infty}^{\infty} d^2 x \  e^{i {\vec q} \cdot {\vec x}} u_{\parallel}({\vec x})$ and ${\vec u}_{\perp}({\vec q}) = \int_{-\infty}^{\infty} d^2 x \  e^{i {\vec q} \cdot {\vec x}} {\vec u}_{\perp}({\vec x})$. Here $\kappa(q)$ and $\mu(q)$, $\lambda(q)$ are the renormalized wave vector dependent bending modulus and elastic moduli, respectively, and we assume the scaling forms
\begin{eqnarray}
\kappa(q) \simeq \kappa_0 (q \xi)^{- \epsilon}
\label{E_q_b}
\end{eqnarray}
\begin{eqnarray}
 \mu(q) \simeq \mu_0 (q \xi)^{\omega}, \   \lambda(q) \simeq \lambda_0 (q \xi)^{\omega}
\label{E_q_in}
\end{eqnarray}
with the exponents $\epsilon$ and $\omega$. In kDNA, the elasticity is entropic in origin, we expect $\kappa_0 \simeq k_BT$, $\mu_0 \simeq \lambda_0 \simeq k_BT/\xi^2 $.

From the free energy~(\ref{f_eff}), one obtains the mean square undulation fluctuation
\begin{eqnarray}
\langle u_{\parallel}^2 \rangle \simeq \xi^2 N^{2 \chi_{\parallel}}
\label{u_pal}
\end{eqnarray}
where the roughness exponent $\chi_{\parallel} = 1 - \epsilon/2$ characterizes how the height fluctuation of the sheet grows with its linear dimension $N$. Similarly, one can calculate the mean square in-plane fluctuation
\begin{eqnarray}
\langle {\vec u}_{\perp}^2 \rangle \simeq  \xi^2 N^{2 \chi_{\perp}}
\label{u_perp}
\end{eqnarray}
with $\chi_{\perp} = \omega/2$.
These exponents have been intensively analyzed in past numerical simulations, which yield $\chi_{\parallel} = 0.6 \sim 0.65$ and $\chi_{\perp} = 0.2 \sim 0.3$, or equivalently, $\epsilon = 0.7 \sim 0.8$ and $\omega = 0.4 \sim 0.6$~\cite{Abraham1989,Nelson_1987_2,Bowick2001,Mizuochi_2014,Chaikin_book}. These values are consistent with the relation $\omega = 2(1-\epsilon)$, which is expected to hold from the rotational invariance of the sheet~\cite{Aronovitz1988,Chaikin_book}.

For the remainder of the paper, we introduce a symbol $\times$ to designate $\parallel$ or $\perp$ for the brevity of the notation. This enables us to rewrite Eqs.~\eqref{u_pal} and~\eqref{u_perp} in a unified form
\begin{eqnarray}
\langle u_{\times}^2 \rangle \simeq \xi^2 N^{2 \chi_{\times}}
\label{u_times}
\end{eqnarray}

\paragraph{Dynamics}
A recent experiment has observed the anomalous dynamics of a quantum dot attached to kDNA. This may be translated in our description to the dynamics of a tagged monomer in the sheet.

Anomalous dynamics due to the two-dimensional connectivity would show up on a time scale longer than the basic time scale $\tau_{\xi}$, see Eq.~\eqref{tau_xi}.
Here, we expect that there are three time scales to characterize the large scale dynamics of the sheet. One is the rotational relaxation time or diffusion time scale $\tau_R$, see Eq.~(\ref{tau_R_1}), i.e., the time needed to diffuse its own size. Others are the conformational relaxation times, one for the out-of-plane undulation mode $\tau_{\parallel}$ and the other for the in-plane phonon mode $\tau_{\perp}$. Unlike flexible polymer case, in which these three time scales degenerate, the sheet in flat phase is expected to exhibit richer dynamics with these time scales well separated. 
In addition, for kDNA, the ``monomer" of size $\xi$ is actually ring, which has internal degrees of freedom. Therefore, one can resolve the shorter time scale $\tau < \tau_{\xi}$ dynamics as well, where the effect of two-dimensional connectivity is irrelevant.

We start our discussion of kDNA dynamics with the shortest time scale $\tau < \tau_{\xi}$. In this regime,  individual rings behave as if they were independent. One thus expects the standard polymer dynamics. Writing the displacement of a tagged bead as $\Delta r$, its the mean-square displacement (MSD) follows~\cite{Panja_2010,Sakaue_2013,Saito_2015,Rubinstein_book} 
\begin{eqnarray}
\langle \Delta r^2(\tau) \rangle \simeq b^2 \left( \frac{\tau}{\tau_{b}} \right)^{\alpha_0}  \qquad (\tau_b < \tau < \tau_{\xi})
\label{MSD_short_0}
\end{eqnarray}
with 
\begin{eqnarray}
\alpha_0 = \left\{
\begin{array}{ll}
 2 \nu/(1+2 \nu) & (\rm{Rouse})\\
  2/3& (\rm{Zimm})
\end{array}
\right.
\label{MSD_short_0_exponent}
\end{eqnarray}
At this point, we recall that the sheet is globally anisotropic with in-plane and out-of-plane directions being well-defined. Anticipating the importance of a clear identification of these directions, we aim to decompose the displacement of the tagged bead into in-plane ($\Delta r_{\perp}$) and out-of-plane ($\Delta r_{\parallel}$) contributions.  
Adopting the component symbol $\times $, see Eq.~\eqref{u_times}, we can rewrite Eq.~\eqref{MSD_short_0} in the component-wise form
\begin{eqnarray}
\langle \Delta r_{\times}^2(\tau) \rangle \simeq c_{\times} b^2 \left( \frac{\tau}{\tau_{b}} \right)^{\alpha_0}  \qquad (\tau_b < \tau < \tau_{\xi})
\end{eqnarray}

Note that the out-of-plane and the in-plane directions share the same exponent $\alpha_0$ due to the isotropy in this short time and length scales, but their amplitudes, represented by the ${\mathcal O(1)}$ coefficients $c_{\parallel}$ and $c_{\perp}$ are different. Reflecting the number of freedoms in each direction, we expect their ratio $c_{\perp}/c_{\parallel} = 2$. This amplitude ratio will be inherited to the longer time scale, and as we shall see below, it plays a role in the dynamical crossover observed in the experiment.

We now turn to the polymerized membrane regime with intermediate time scale $\tau_{\xi} < \tau$, and closely follow the dynamical scaling argument in ref~\cite{Mizuochi_2014}.
First, recall the center-of-mass diffusivity $D_{cm}\sim N^{-2}$ for Rouse dynamics and  $D_{cm}\sim N^{-1}$ for Zimm dymamics, see Eq.~(\ref{D_cm}).
For each of out-of-plane undulation mode and in-plane phonon mode, we obtain the relaxation time 
$\tau_{\times}$ from the relation $D_{cm} \tau_{\times} \simeq \langle u_{\times}^2 \rangle$;
\begin{eqnarray}
\tau_{\times} \simeq \left\{
\begin{array}{ll}
\tau_{\xi}N^{2 \chi_{\times}+2} & (\rm{Rouse})\\
\tau_{\xi}N^{2 \chi_{\times}+1} & (\rm{Zimm})
\label{t_times}
\end{array}
\right.
\end{eqnarray}
Assuming a self-similar dynamics for the displacement $\Delta r_{\times}$ of a labeled ring (or a tagged bead in the ring) in each direction for time scale $\tau_{\xi} < \tau < \tau_{\times}$
\begin{eqnarray}
\langle \Delta r_{\times}^2(\tau) \rangle \simeq  c_{\times} \xi^2 \left( \frac{\tau}{\tau_{\xi}} \right)^{\alpha_{\times}} \qquad (\tau_{\xi} < \tau < \tau_{\times})
\label{MSD_times_1}
\end{eqnarray}
and requiring the matching condition $\langle \Delta r_{\times}^2(\tau) \rangle  \xrightarrow{\tau \rightarrow \tau_{\times}} \langle u_{\times}^2 \rangle$, we find
\begin{eqnarray}
\alpha_{\times} = \left\{
\begin{array}{ll}
\frac{\chi_{\times}}{\chi_{\times}+1}  & (\rm{Rouse})\\
\frac{2\chi_{\times}}{2\chi_{\times}+1} & (\rm{Zimm})
\end{array}
\right.
\end{eqnarray}
Substituting the reported values for the roughness exponents $\chi_{\parallel}$ and $\chi_{\perp}$, we thus obtain the following set of growth exponents: $(\alpha_{\parallel}, \alpha_{\perp}) \simeq (0.37, 0.2)$ for Rouse dynamics, and  $(\alpha_{\parallel}, \alpha_{\perp}) \simeq (0.55, 0.3)$ for Zimm dynamics.

On longer time scale $\tau > \tau_{\times}$, all the internal modes in the respective direction ($\parallel$ or $\perp$) relax, hence, the diffusive dynamics of the center-of-mass is expected;
\begin{eqnarray}
\langle \Delta r_{\times}^2(\tau) \rangle \simeq c_{\times}  \langle u_{\times}^2 \rangle \left( \frac{\tau}{\tau_{\times}} \right) \qquad (\tau_{\times} < \tau)
\end{eqnarray}
We note that since $\tau_R \gg \tau_{\parallel}, \tau_{\perp}$, our description assuming distinctive directions for the internal relaxation dynamics makes sense.
One can also check the expected crossover $\langle \Delta r_{\times}^2(\tau) \rangle \xrightarrow{\tau \rightarrow \tau_{R}} R^2$, where $R \simeq \xi N$, see the flat state scaling~\eqref{R_flat}. At this time scale, the rotational relaxation takes place, and the dynamics becomes isotropic on even longer time scale.

\section{Discussions}
In this section, we compare our prediction on kDNA dynamics with recent experimental observations, moslty referring to Refs.~\cite{Klotz_2019,Ramakrishnan_2025}. We also give an estimate for the kDNA elasticity based on the polymerized membrane model and compare with the estaimte in Ref.~\cite{He2023}.
\paragraph{Dynamics}
To obtain the order of magnitude estimation for characteristic time scales of kDNA, we need reasonable values for the kDNA size and the basic time scale. First, from the observation that kDNA consists of thousands of minicircles, we assume $N \approx  50 \sim 100$ for the typical linear dimension of kDNA sheet. Second, the ``monomers" of kDNA is a DNA ring with its spatial size $\xi \sim 0.1$ ($\mu$m)~\cite{He2023}, the corresponding time scale may be estimated as $\tau_{\xi} \approx 1 \sim 10$ (ms) in aqueous environment.  Given Eq.~\eqref{u_pal} and the roughness exponent $\chi_{\parallel}= 0.6 \sim 0.65$, the measured undulation amplitude $\langle u_{\parallel} ^2 \rangle \approx 1$ ($\mu {\mathrm m}^2$) is consistent with the above estimate for $N$ and $\xi$.

Using these estimates, we obtain from Eq.~(\ref{tau_R_1})
\begin{eqnarray}
\tau_R \approx \left\{
\begin{array}{ll}
 10^{4}  \sim 10^6\ (\rm{s})  &(\rm{Rouse})\\
10^2  \sim 10^4  \ (\rm{s}) & (\rm{Zimm})
\end{array}
\right.
\end{eqnarray}
and from Eqs.~(\ref{t_times})
\begin{eqnarray}
\tau_{\parallel} \approx \left\{
\begin{array}{ll}
  10^{2} \sim  10^4  \ (\rm{s}) & (\rm{Rouse})\\
1 \sim 10^2  \ (\rm{s}) & (\rm{Zimm})
\label{t_parallel_es}
\end{array}
\right.
\end{eqnarray}
\begin{eqnarray}
\tau_{\perp} \approx \left\{
\begin{array}{ll}
10  \sim  10^3  \ (\rm{s}) & (\rm{Rouse})\\
0.1 \sim 10  \ (\rm{s}) & (\rm{Zimm})
\label{t_perp_es}
\end{array}
\right.
\end{eqnarray}
We find that the predictions based on Zimm dynamics well match the experimental observations in Ref.~\cite{Klotz_2019}: $\tau_R \sim $  several minutes; and two time scales $\tau_{fast} \sim 0.2 (s)$ and $\tau_{slow} \sim 6.5 (s)$ found in the anisotropy autocorrelation.

\begin{figure*}[t]
	\centering
\includegraphics[width=0.95\textwidth]{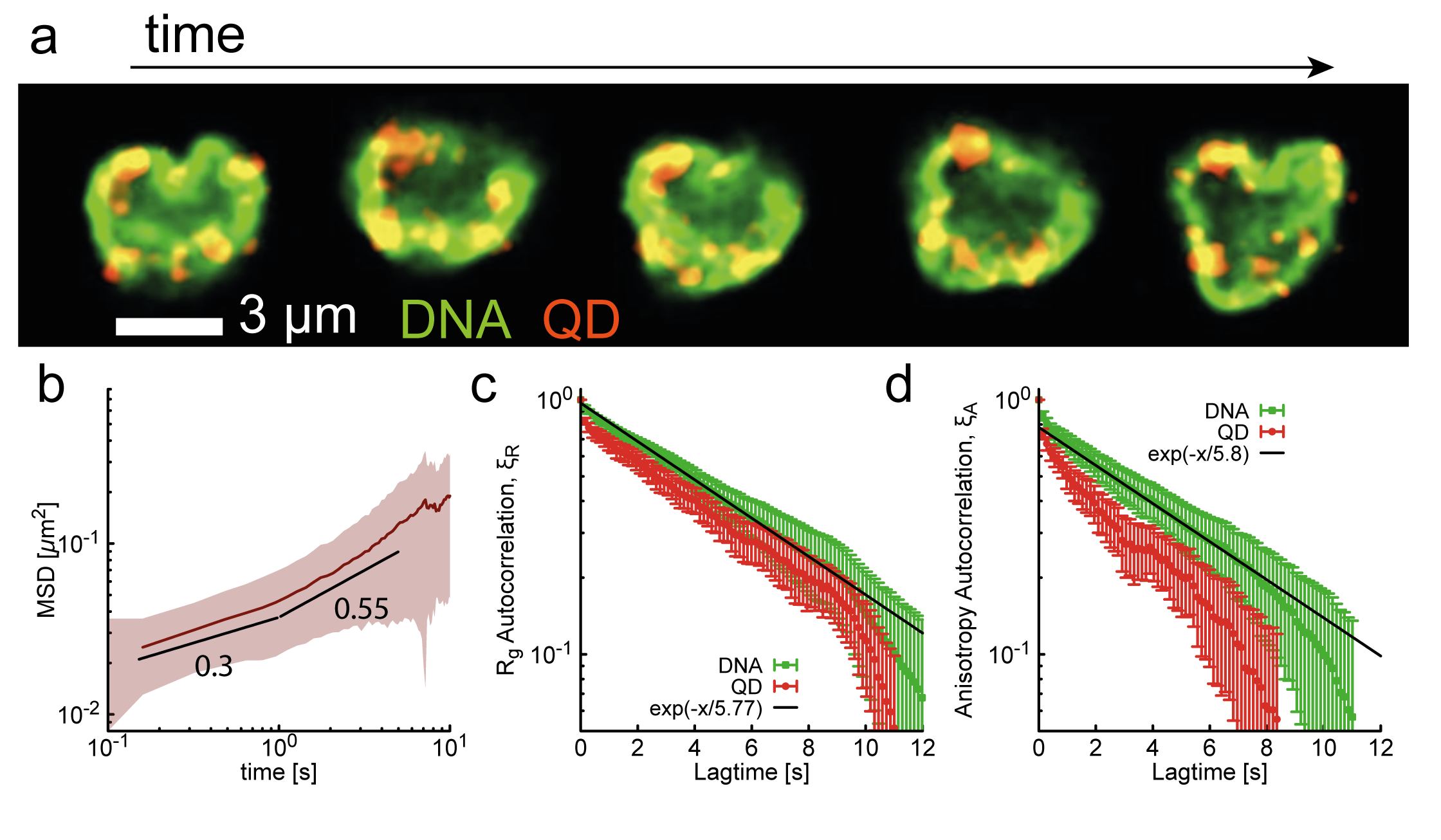}
\vspace{-0.5 cm}
	\caption{ \textbf{Dynamics of kDNA}. \textbf{a.} Snapshots taken from a video of a kDNA diffusing in aqueous solution; the DNA (green) is labelled with YOYO-I intercalator while quantum-dots (red) are bound to specific DNA sequences in the kDNA. \textbf{b.} Mean Square Displacement (MSD) calculated in the center of mass frame of quantum dots attached to kDNA. Slopes with exponents $0.3$ and $0.55$ are shown in black as guides for the eye. \textbf{c.} Autocorrelation of the radius of gyration $\xi_R$ computed from the DNA signal (green) or the quantum dots signal (red). The black line follows an exponential with characteristic decay time $\tau_R \simeq 5.77$ seconds. \textbf{d.} Autocorrelation of the anisotropy $\xi_A$ computed from the DNA signal (green) or the quantum dots signal (red). The black line follows an exponential with characteristic decay time $\tau_R \simeq 5.8$ seconds. Data are taken from Ref.~\cite{Ramakrishnan_2025}.}
	\label{Fig2}
		\vspace{-0.4 cm}
\end{figure*}

Figure~\ref{Fig2} shows an example of the experimentally observed dynamics of kDNA in aqueous solution, where several quantum dots (red) are attached to specific DNA sequences within the kDNA structure (green) to track its internal dynamics~\cite{Ramakrishnan_2025}.
From the time lapse images of these quantum dots, we calculated the mean square displacement (MSD) of the QDs in their center of mass frame, and plotted it as a function of lagtime in Fig.~\ref{Fig2}(b).
Because of the factor $c_{\perp}/c_{\parallel}$, we expect that the in-plane displacement initially dominates the MSD, but the out-of-plane displacement with the larger growth exponent eventually overtakes. This leads to the crossover kinetics for the observed MSD, where the early and the late stages are characterized by the growth exponents $\alpha_{\perp}$ and $\alpha_{\parallel}$, respectively, on the time scale $\tau_{\xi} < \tau <  \tau_{\parallel}$.
Again, the prediction from Zimm dynamics $\alpha_{\perp} \simeq 0.3$ and $\alpha_{\parallel} \simeq 0.55$ agrees with the experimental observation~\cite{Ramakrishnan_2025}.
We also calculated the autocorrelation function of the radius of gyration (Fig.~\ref{Fig2} (c)) and the shape anisotropy (Fig.~\ref{Fig2} (d)) of kDNA from the DNA signal (green) and the quantum dots signal (red). From the single exponential fit, we deduced a relaxation time $\simeq 5.8$ (sec) that likely corresponds to $\tau_{slow}$ measured in the previous experiment~\cite{Klotz_2019}.

\paragraph{Elastic response}
kDNAs are expected to exhibit a very soft elastic response for two reasons. First, viewing kDNA as a two-dimensional lattice, its lattice constant $\xi$ (monomer size) is much larger than the atom size. Second, two-dimensional systems are generally subject to long-wavelength fluctuations at finite temperature. In a simple two-dimensional lattice with only in-plane phonon mode, this leads to a logarithmic softening of the elastic modulus with the system size. For kDNA, the presence of the out-of-plane undulation mode and its nonlinear coupling with phonon mode leads to an anomalous elasticity, where the in-plane elasticity softens on larger scale according to a power-law. Note that the concurrent stiffening in the bending rigidity stabilizes the flat phase. Such features are represented in the free energy functional~(\ref{f_eff}) via the wave number dependent bending and elastic moduli, see Eqs.~(\ref{E_q_b}) and ~(\ref{E_q_in}).

The overall in-plane elasticity of kDNA is thus estimated as
\begin{eqnarray}
\mu(q_{min}) \simeq \lambda (q_{min}) \simeq \frac{k_BT}{\xi^2 N^{\omega}} 
\label{mu_lambda_global}
\end{eqnarray}
where $q_{min} \simeq (\xi N)^{-1}$ is the smallest wave number. With $N = 50 \sim 100$ and $\omega = 0.4 \sim 0.6$, we obtain an estimate $\mu(q_{min}) \sim \lambda (q_{min}) \sim 10$ in unit of $k_BT/ \mu m^2$.
We note that by combining Eqs.~\eqref{u_pal} and ~\eqref{mu_lambda_global}, the elastic modulus can also be evaluated from the measured undulation amplitude
\begin{eqnarray}
\mu(q_{min}) \simeq \lambda (q_{min})  \sim \frac{k_BT}{\xi^2} \left[ \frac{\langle u_{\parallel}^2\rangle}{\xi^2}\right]^{-\omega/(2 \chi_{\parallel})}
\end{eqnarray}

The calculatin for $\mu(q_{min}) \sim \lambda (q_{min}) \sim 10 k_BT/ \mu m^2$ is equivalent to 0.04 pN/$\mu$m, which is remarkably close to the experimentally estimated stretch modulus $Y \simeq 0.1$ pN/$\mu$m in Ref.~\cite{He2023}. These values are about $10^6$ times smaller than those of lipid bilayers, which display large stretch moduli, $Y \simeq 0.1\sim 1$ N/m~\cite{JANSHOFF20152977}. These extremely small values of elasticity are physically justified by the fact that each ``monomer'' in the kDNA network is very large and displays significant conformational freedom, effectively translating into very weak springs in the polymerised membrane model.

\section{Summary}
kDNAs found in the mitocondorion of kinetoplastid organisms have long been known to take a topologically complex architecture with thousands of DNA rings. Despite recent growing interest on the kDNA from a physical and material science point of view, there is no consensus yet on how to describe its material properties.  In this paper, we have suggested that the large scale behaviors of the kDNAs are subject to a strong geometrical constraint due to the two-dimensional catenated network of component rings, hence, their various unique properties could be understood based on the physics of the polymerized membrane. We have demonstrated that this view indeed explains the results on the static and dynamics of kDNA revealed in recent imaging experiments in a consistent way.

Nevertheless, we should not forget the complexity of the real kDNA; it is not merely a connected network of monodisperse rings, but is rather often made as a bidisperse mixture of rings with different sizes, i.e., mini- and maxi-circles~\cite{Michieletto2025nar}. The success of our simple description may indicate the identification of the polymerized membrane as a base model of kDNAs that allows us to describe some essential aspects of the kDNA behaviors. It would be an interesting challenge to analyze the rich phenomenology caused by, for instance, the introduction of maxicircles by extending the current simple description.

{\sl  Acknowledgments} --
This work is supported by JSPS KAKENHI (Grant No. JP23H00369 and JP24K00602).

\section{Appendix}
In the main text, we have seen that the kDNA modeled as a polymerized membrane does not crumple, rather takes a flat conformation. Nevertheless, it may be interesting to see how the hypothetical crumpled structure behaves. In this appendix, we briefly outline the dynamics of the crumpled sheet. 

We recall that the basic time scale of the sheet~\eqref{tau_xi} and the diffusivity of the center of mass of the sheet~\eqref{D_cm}. Then, from the relation $D_{cm} \tau_R \simeq R^2$, we get the relaxation time of the sheet;
\begin{eqnarray}
\tau_R \simeq \left\{
\begin{array}{ll}
\tau_{\xi}M^{9/5} & (\rm{Rouse})\\
\tau_{\xi}M^{6/5} & (\rm{Zimm})
\label{t_r_crumpled}
\end{array}
\right.
\end{eqnarray}

For the MSD of a labeled bead, we expect a standard polymer scenario, see Eqs.~\eqref{MSD_short_0} and~\eqref{MSD_short_0_exponent} up to the time scale $\tau_{\xi}$.
On longer time scale $\tau > \tau_{\xi}$, where two-dimensional connectivity set in, we may assume the following scaling form
\begin{eqnarray}
\langle \Delta r^2(\tau) \rangle \simeq \xi^2 \left( \frac{\tau}{\tau_{\xi}} \right)^{\alpha}
\end{eqnarray}
and we expect that this regime persists up to the time scale $\tau_R$. Requiring $\langle \Delta r^2(\tau) \rangle  \rightarrow R^2$ ($\tau \rightarrow \tau_R$), we find
\begin{eqnarray}
\alpha = \left\{
\begin{array}{ll}
4/9 & (\rm{Rouse})\\
2/3& (\rm{Zimm})
\end{array}
\right.
\end{eqnarray}

\providecommand{\noopsort}[1]{}\providecommand{\singleletter}[1]{#1}%

\end{document}